# Efficient continuous-wave nonlinear frequency conversion in high-Q Gallium Nitride photonic crystal cavities on Silicon

Mohamed Sabry Mohamed[1,*], Angelica Simbula[2], Jean-François Carlin[1], Momchil Minkov[1,3], Dario Gerace[2], Vincenzo Savona[1], Nicolas Grandjean[1], Matteo Galli[2], and Romuald Houdré[1]

[1]Institute of Physics, Ecole Polytechnique Fédérale de Lausanne (EPFL), Switzerland
[2]Dipartimento di Fisica "A. Volta," Università di Pavia, via Bassi 6, 27100 Pavia, Italy
[3]Currently with Ginzton Laboratory, Stanford University, Stanford, CA 94305, USA
*Corresponding author: mohamedsabry.mohamed@epfl.ch

We report on nonlinear frequency conversion from the telecom range via second harmonic generation (SHG) and third harmonic generation (THG) in suspended gallium nitride slab photonic crystal (PhC) cavities on silicon, under continuous-wave resonant excitation. Optimized two-dimensional PhC cavities with augmented far-field coupling have been characterized with quality factors as high as $4.4 \times 10^4$, approaching the computed theoretical values. The strong enhancement in light confinement has enabled efficient SHG, achieving normalized conversion efficiency of $2.4 \times 10^{-3}$ W$^{-1}$, as well as simultaneous THG. SHG emission power of up to 0.74 nW has been detected without saturation. The results herein validate the suitability of gallium nitride for integrated nonlinear optical processing.

Photonic crystal (PhC) cavities offer a viable mechanism for light localization, particularly in planar geometries, which have potential for monolithic integration. While high refractive index semiconductors such as silicon (Si) have been thoroughly investigated due to their enhanced confinement capability, their optical response in the near-infrared (near-IR) range is typically dominated by undesirable nonlinear effects when sufficiently high field strength is reached within the dielectric material, which is the case encountered in high quality (Q)-factor resonant microcavities. Such effects may range from optical bistability to peak-dragging and linewidth broadening [1,2], induced by two-photon absorption, free-carrier absorption, and thermal instability. Fortunately, the aforementioned limitations can be overcome in wide bandgap semiconductors. Gallium nitride (GaN) for instance possesses a bandgap that extends to the near-UV, which accommodates the optical transitions of many well-behaved emitters and offers the transparency needed for multi-photon processes, allowing for efficient nonlinear optical interactions. Nonlinear optical effects are particularly appealing for integrated optics, as photons can interact indirectly to provide light control modalities that include: optical switching, phase control, and frequency conversion. The strong second-order susceptibility of the non-centrosymmetric wurtzite crystal structure of GaN enables a variety of second-order nonlinear effects, which can be harnessed with a small footprint through the use of PhCs that appropriately mediate light-matter interaction. Additional enhancement can be introduced by high-Q cavities, such that nonlinearity is achieved with low photon count. Following the development of the GaN-on-Si platform [3] and further efforts that yielded improved figures for cavity Q-factors [4,5], resonant harmonic generation has been recently demonstrated in GaN PhC cavities [6,7], although with limited efficiency.

In this report, we present GaN slab PhC cavities on Si operating in the telecom range, with optimized designs where the fundamental mode is coupled to the far-field. Characterization of fabricated cavities revealed record Q-factors, which has enabled enhanced second harmonic generation (SHG) and third harmonic generation (THG) by means of resonant far-field excitation. High conversion efficiencies have been achieved in GaN under continuous-wave operation. Factors contributing to the observed enhancement in nonlinear conversion are highlighted within the context of PhC design.

Two cavity designs were considered for light localization in a two-dimensional, triangular PhC lattice: the L3 cavity, based on three missing holes [8], and the H0 (point-shift) cavity, which displays the lowest modal volume amongst planar, two-dimensional PhC cavities [9]. To allow for the manipulation of the out-of-plane radiative losses, the first three side holes of the L3 cavity were modified both in size and position as indicated in **Fig. 1**, while with the H0 cavity, the first five side holes and two vertical holes were adjusted in position. Two design wavelengths in the telecom range (at λ=1300 nm and λ=1550 nm) were targeted for each cavity by scaling the lattice constant (**a** = 470 nm and 570 nm respectively) while maintaining a lattice-constant to hole-radius ratio of 4:1, for a fixed slab thickness of 350 nm.

An optimization algorithm based on guided-mode expansion [10,11] was utilized to sweep parameter space within specified constraints for the highest attainable Q-factors. The optimization runs yielded cavities with a theoretical Q-factor of $1.1 \times 10^5$ for both L3 and H0 cavities resonant near λ=1300 nm, while Q-factors of $2.2 \times 10^5$ and $1.2 \times 10^5$ were obtained for L3 and H0 cavities respectively, resonant near λ=1550 nm. The higher Q-factors obtained at λ=1550 nm were in part due to the chosen slab thickness. Simulations using finite-difference time-domain (FDTD) and finite element method solvers confirmed the computed profiles and Q-factors of the cavity modes.

Further modification of the PhC lattice served to enhance coupling to the fundamental cavity mode through band folding. The PhC holes surrounding the cavity as shown in **Fig. 1a**, referred to henceforth as "injectors", were increased in radius relative to the native PhC holes such that a twofold periodicity (2a) is imposed on the lattice. This acts to fold the k-vector components present at the Brillouin zone edge to the Γ-point in reciprocal space [12], thus increasing the overlap with the excitation aperture. The resultant far-field emission pattern of the cavities is displayed in **Fig. 1b** for an injector-hole radius incrementation ($\Delta r_i$) of 10 nm. Contrary to the near-Gaussian distribution of the far-field pattern of the L3 cavity, two lobes appear around the center of reciprocal space for the H0 cavity, arising from the odd parity of the fundamental mode. Cavity Q-factors gradually dropped with increasing injector-hole radius [13], down to a value of 2.7–4.4×$10^4$ for the largest considered injector sizes ($\Delta r_i$=8 nm and 10 nm for the L3 and H0 cavities, respectively).

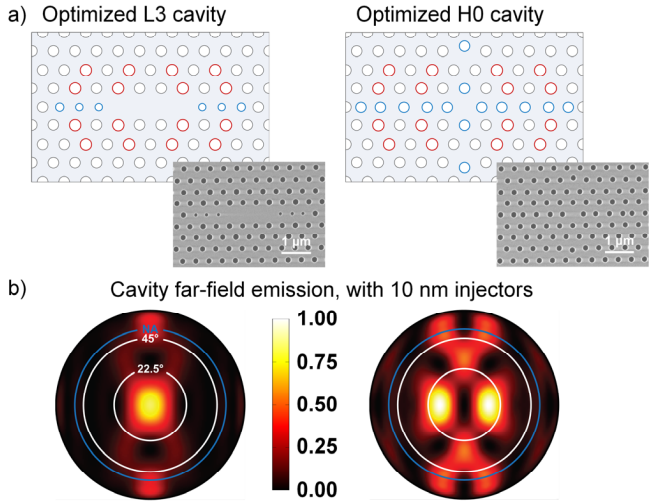

**Fig. 1.** (a) Schematic diagrams of the implemented L3 and H0 optimized cavity designs. The holes highlighted in blue are modified in size for the L3 cavity and in position for both cavities, while the holes highlighted in red are enlarged relative to the native lattice to impart band folding. Scanning electron micrographs of the fabricated GaN L3 and H0 cavities, resonant near 1300 nm, are shown in the inset. (b) The far-field radiative-intensity ($|\mathbf{E}|^2$) light cone for each of the L3 and H0 cavity designs respectively, with a/λ=0.37, for $\Delta r_i$=10 nm. The circles from inside out outline the following emission angles: 22.5°, 45°, and a numerical aperture (NA) of 0.8 (in blue).

We have selected three injector sizes to be fabricated for each cavity, ranging in $\Delta r_i$ between 3 nm and 10 nm ($\Delta r_i$=[3, 5, 8] nm for the L3 cavity and $\Delta r_i$=[5, 7, 10] nm for the H0 cavity) to cover a coupling efficiency (η) range from approximately 2% to 20%. The PhC cell was chosen to be 65**a** by 39√3**a** units in size, with the cavity residing at the center. The sample design accounted for lithographic tuning of the PhC-hole radius around the nominal optimization values to target the design wavelength.

Heteroepitaxial growth of GaN was carried out by metal-organic chemical vapor deposition on a Si (111) wafer. A nucleation layer consisting of 40 nm of aluminum nitride (AlN) was first deposited, followed by 310 nm of GaN. The grown gallium-face layer is c-axis oriented, with similar epitaxial properties to previous reports [3].

The fabrication of the PhCs was initiated by depositing a $SiO_2$ layer using plasma-enhanced chemical vapor deposition, to function as a hard etch mask. ZEP520A positive resist was spun on the $SiO_2$ layer, then patterned by a 100 keV electron beam (Vistec EBPG5000). After development, the pattern was transferred to the $SiO_2$ layer by inductively-coupled plasma reactive-ion etching (ICP-RIE). The hard mask was then used to etch the III-nitride layer by ICP-RIE using chlorine-based chemistry, and was removed afterwards by hydrofluoric acid etching. The GaN membrane was finally released by $XeF_2$ vapor-phase etching of Si.

To characterize the fabricated PhC cavities, we employed resonant scattering, which has been shown to be a rapid and reliable technique to extract cavity Q-factors without introducing additional coupling channels [14,15]. For that configuration, the cavity axis is oriented at 45° relative to a vertical polarizer, which allows for exciting both $E_x$ and $E_y$ components of the cavity modes under far-field illumination. The back-reflected signal is filtered in cross polarization, by a horizontal analyzer. Careful adjustment of the coupling should maximize the resonantly scattered signal of the mode over the background from the PhC lattice to produce a symmetric Lorentzian line shape.

Near the design wavelength of 1550 nm, Q-factors above 2.5×$10^4$ were measured for the investigated range of injector sizes. Record Q-factors as high as 4.1×$10^4$ and 4.4×$10^4$ were observed with L3 and H0 (displayed in **Fig. 2**) cavities respectively, both of which exhibited intermediate-sized injectors (5 nm and 7 nm respectively) and a theoretical Q-factor ≈8.0×$10^4$. The measured Q-factors of their 1300 nm counterparts were marginally lower on average, indicating that scattering and disorder-induced losses are the dominant mechanisms setting the upper limit on light confinement. A common expected trend was the notable degradation in Q-factor, both away from the optimized-design wavelength and with increasing $\Delta r_i$. A spread in the Q-factor and cavity resonance wavelength amongst replicas was found, similar to earlier reports [5], as a consequence of the inhomogeneity of the GaN-AlN interface and residual fabrication disorder.

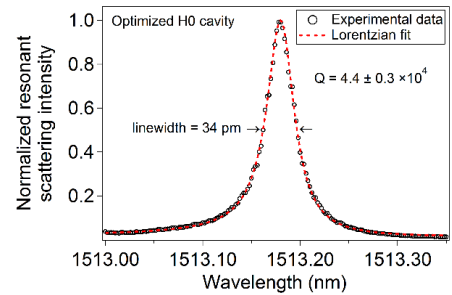

**Fig. 2.** Resonant scattering spectrum of the fundamental mode of an optimized H0 cavity at λ=1513 nm, exhibiting a linewidth of 34 pm, which corresponds to a Q-factor of 4.4×$10^4$.

We have utilized resonant scattering to further estimate the coupling efficiency to the fundamental cavity mode. After measuring the resonant scattering signal using the standard cross-polarization configuration, power impinging on the cavity was evaluated in parallel-polarization (analyzer relative to the polarizer), after replacing the sample with a mirror. The ratio of the resonantly scattered light intensity to the total incident intensity is indicative of the fraction of light that couples into the cavity. For the fabricated range of injector sizes, the experimental coupling efficiencies at λ=1550 nm were measured to be between 3% and 19% for L3 cavities, and between 2% and 15% for H0 cavities, confirming the expected dependence.

Next, harmonic generation from the PhC cavities was investigated. A near-IR, tunable continuous-wave laser source was used to irradiate the sample with a linearly polarized beam, oriented such that the electric field is perpendicular to the horizontal axis of the PhC cavity. The collimated laser beam was tightly focused at the sample plane by a microscope objective (50x, 0.8 NA), and coupling to the cavity was optimized through fine translation of the sample in the transverse plane and axial direction using a piezo stage. The upconverted signal was collected through the same objective and redirected for detection by a dichroic mirror (detailed configuration can be found in [16]).

When excited on resonance, second harmonic generation from PhC cavities was directly evident through a relatively strong and distinct emission pattern emanating from the center. **Fig. 3** displays the observed SHG emission at the sample plane from a L3 and a H0 cavity, resonantly excited at λ=1545 nm and λ=1513 nm, respectively. The z-component of the second-order polarization is shown in the inset for comparison. It was computed by considering the electric field of the fundamental mode, using the following expression:

$$P_z^{(2\omega)} = \frac{\varepsilon_0}{2}\left(\chi_{zxx}^{(2)}E_x^{(\omega)^2} + \chi_{zyy}^{(2)}E_y^{(\omega)^2} + \chi_{zzz}^{(2)}E_z^{(\omega)^2}\right), \quad (1)$$

where $\chi_{zxx}^{(2)} = \chi_{zyy}^{(2)} = -0.5\chi_{zzz}^{(2)}$ [17]. The dominant components of the second-order susceptibility tensor of wurtzite GaN ($C_{6v}$ point-group symmetry) are $\chi_{zzz}^{(2)}$ followed by $\chi_{zxx/zyy}^{(2)}$ and $\chi_{xxz/yyz}^{(2)}$ (and their respective permutations) [18], of which $\chi_{zxx/zyy}^{(2)}$ can be exploited with the quasi transverse-electric nature of the fundamental cavity mode, for the c-axis oriented GaN layer. These tensor components act to couple the in-plane electric field ($E_{x/y}$) to the second-order polarization ($P_z^{(2\omega)}$) along the z-axis. The induced polarization hence follows the spatial profile of the resonant fundamental mode.

However, it is evident that the imaged SHG emission pattern does not directly map the cavity mode. This is due to the transparency of the GaN layer that is multi-modal at the harmonic wavelength, combined with the symmetry of the $\chi^{(2)}$ tensor. The generated harmonic originating from the bulk nonlinearity is propagating and appears to be strongly scattered upon interaction with the PhC lattice. The extended parallel lines previously reported in SHG experiments are present in both cavities [6,19], along with the cross-shaped pattern, where the second harmonic is seen to traverse the lattice along the ΓK direction. With the H0 cavity, a ring appears at the center, overlapping with the position of the first set of holes surrounding the polarization hotspots. Modeling of SHG emission by FDTD has confirmed guided propagation of the SHG signal in the GaN slab between PhC holes, primarily along the diagonal directions. Scattering occurs adjacent to the PhC-hole boundaries where the propagation of the generated harmonic is obstructed. It should be noted that the PhC lattice does not exhibit any bandgap at the SHG wavelength, and consequently, no doubly-resonant enhancement effects are expected.

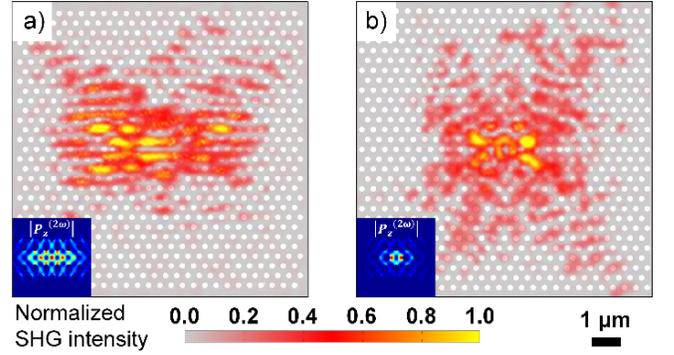

**Fig. 3.** SHG spatial emission profile of an optimized (a) L3 and (b) H0 cavity, overlaid on a schematic of their respective PhC lattice, with the normalized SHG intensity indicated in false color, and the z-component of the second-order polarization ($|P_z^{(2\omega)}|$) in the inset.

The nature of the second harmonic process was verified through power dependence measurements. A Si photodetector was attached to the output port after the dichroic mirror, behind an optical filter that suppressed the source laser wavelength. Off-resonance, no power was detected from the PhC cavities across the spectrum under the highest incident power of 6 mW. The collected SHG power was then measured directly under resonant excitation and is shown in **Fig. 4a** as a function of the power coupled into the L3 cavity. This cavity exhibits a coupling efficiency of 13% and a Q-factor of $3.3\times10^4$. A fit to the data gives a slope of 2.0, confirming the expected quadratic dependence. The characteristic Lorentzian-squared spectral profile of SHG was captured and is displayed in **Fig. 4b**, alongside the resonant scattering spectrum for comparison. The relative blue-shift (Δλ ≈ 9 pm) of SHG, seen in the plot, was irreversible and is hypothesized to be a consequence of photo-induced oxidation.

Under optimal excitation conditions at resonance, the collected SHG power from the L3 cavity reached 0.74 nW for 0.78 mW of power coupled into the cavity. Accounting for SHG emission into full-space by a multiplication factor of two, this corresponds to a normalized conversion efficiency ($P_{SHG}/P_{coupled}^2$) of $2.4\times10^{-3}$ W$^{-1}$. This value is a lower bound on the efficiency of the SHG process since the collection of the SHG signal is limited by the numerical aperture of the objective, and furthermore, we do not account for the propagating portion of the generated harmonic. Similar SHG power was obtained with a L3 cavity resonant at λ=1309 nm, which reached 0.20 nW for 0.43 mW of coupled power, achieving a conversion efficiency of $2.2\times10^{-3}$ W$^{-1}$ (not shown). This confirms performance scalability in the telecom window. Compared to L3 cavities, SHG collected from the H0 cavities did not reach the same intensity levels. The highest measured

SHG power was 0.13 nW for a cavity resonant at λ=1513 nm. Similar analysis—accounting for 0.63 mW of coupled power—indicates a conversion efficiency of $0.65\times10^{-3}$ W$^{-1}$. The conversion efficiency obtained here exceeds previous reports of SHG in a PhC cavity by more than two orders of magnitude [6], as well as reports of SHG in a ring resonator platform [20]. A compelling aspect is that no saturation effects for the SHG process have been observed for the aforementioned magnitude of injected power, which was at the limit of the employed laser source.

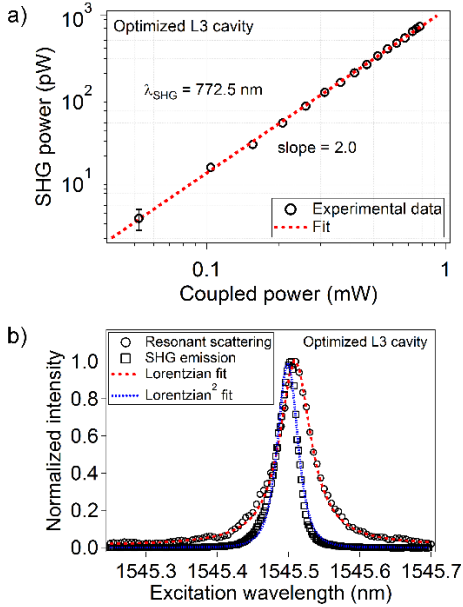

**Fig. 4.** For the optimized L3 cavity, (a) the dependence of the collected SHG power on the coupled excitation power at resonance (λ=1545 nm) and (b) normalized intensity of resonant scattering and SHG emission as a function of source excitation wavelength.

THG was further detected under continuous-wave resonant excitation, with an expected weaker intensity following the third-order nonlinear susceptibility, contrary to similar experiments on PhC cavities in centrosymmetric crystals, such as Si [16]. Amongst the relevant tensor elements of GaN with regard to the cavity mode and crystallographic orientation, the ones with the principal contribution are $\chi^{(3)}_{xxxx}$ and $\chi^{(3)}_{yyyy}$ [18]. This direct dependence on the electric-field components has been recently utilized to map the fundamental cavity mode through THG [7]. For the L3 cavity shown earlier, the emission was fiber-coupled at the output port to a single-mode fiber and sent to a monochromator (600 lines/mm grating) with a cooled Si detector to monitor the intensity count. The spectrum was acquired and is displayed in inset of **Fig. 5**. SHG and THG peaks are seen at 772.5 nm and 515 nm respectively, both below the GaN bandgap (3.4 eV). The ratio between the second and third harmonic intensities was found to be sensitive to the working distance of the objective during the measurement, due to the difference between the radiation patterns of the two harmonic signals. The displayed spectrum has been captured under optimal THG collection. The plot in **Fig. 5** shows the growth ratio of THG intensity relative to SHG as a function of input power. An exponential ratio of 1.4 is extracted, slightly off the expected 3:2 ratio. THG power was estimated to be on the order of 100 fW under incident power of 4.4 mW, based on calibrated power conversion.

The figure of merit for nonlinear frequency conversion in PhC cavities can be formulated by considering the factors contributing to field enhancement at the fundamental frequency, which in turn reduces the threshold power for the nonlinear interaction. As the electric field scales linearly with the cavity Q-factor, coupling efficiency, and inverse of the modal volume (V), an n-order process with n participating degenerate fields thus benefits, ideally, from an enhancement factor of $(\eta Q/V)^n$. The crucial role of the designed injectors is to raise η without significant reduction in the *experimental* Q-factor, boosting the conversion efficiency. The stronger SHG emission observed with L3 cavities relative to H0 cavities of comparable Q-factor, despite possessing a larger modal volume, can be attributed to the enhanced coupling arising from better overlap with the Gaussian excitation beam and the orientation of the mode field components along the y-polarized excitation axis. Further considerations regarding material dependence and radiation of the harmonic signal are left for future investigation.

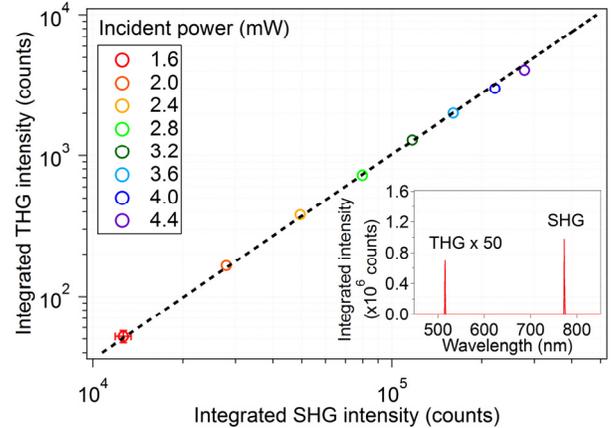

**Fig. 5.** Plot of the collected SHG vs. THG intensity as a function of power incident on an optimized L3 cavity, under resonant excitation at λ=1545 nm. The cavity's emission spectrum is shown in the inset, displaying SHG and THG at a wavelength of 772.5 nm and 515 nm respectively, with the latter scaled for visibility.

In conclusion, we have achieved high nonlinear frequency upconversion efficiency in GaN on Si (of up to $2.4\times10^{-3}$ W$^{-1}$ for SHG) by means of enhanced field coupling and confinement in optimized PhC cavities, which has enabled second and third harmonic generation under continuous-wave, resonant excitation. We have also demonstrated experimental Q-factors as high as $4.4\times10^4$ in the telecom range for GaN PhC cavities that exhibit both high Q/V ratio and far-field excitation capacity. While there is a clear trade-off theoretically between coupling efficiency and Q-factor for a given cavity design, the upper limit on the Q-factor that is imposed by loss channels, given the disorder figure of current fabrication technology, makes room for introducing improved far-field coupling to enhance nonlinear processes without sacrificing the experimentally achievable light confinement. The platform's capability for integration and the fact that the demonstrated nonlinearity has been achieved under continuous-wave operation is promising for on-chip low-power applications, particularly integrated optical networks that can benefit from the generation of non-classical light states, frequency upconversion for improved near-IR detection, and all-optical switching to support emerging optical computing architectures.

The authors would like to acknowledge funding from the Swiss National Science Foundation under projects 200021_146998, 200020_162657, and 200020_149537.